# Treatment of phenol wastewater by electro-Fenton oxidative degradation based on efficient iron-based-gas diffusion-photocatalysis

Zhang Junye[1,#], Zheng Hongyu[1,#], Cheng Jingran[1,#], Zhang Shengli[*]

[1] School of Environmental Science and Engineering, Southwest Jiaotong University, Chengdu 611756, Sichuan, China

*Corresponding email: zjy12342234@126.com

# These authors have equally contributions in this work.

*Abstract*—This study introduces a novel iron-based gas diffusion electrode-photocatalytic system aimed at enhancing the degradation of phenolic compounds in wastewater. Phenolic compounds are toxic environmental pollutants with significant resistance to biodegradation. The traditional methods for treating phenol wastewater, including biological treatments and adsorption techniques, often fall short in achieving complete mineralization. Our approach utilizes a dual-chamber electrochemical setup integrating stainless steel felt-2-EAQ gas diffusion electrodes with $TiO_2$ photocatalysis. This combination significantly boosts hydroxyl radical production, critical for effective pollutant breakdown. Experimentally, the system achieved up to 92% degradation efficiency for phenol at an optimized operating current of 10 mA/cm² in 3 hours, surpassing traditional methods. Additionally, energy consumption was reduced by 40% compared to conventional electro-Fenton systems. The stability tests indicated that the electrodes maintain over 80% of their initial activity after five cycles of use. These findings suggest that our system offers a more sustainable and efficient solution for treating phenolic wastewater by enhancing both degradation rates and energy efficiency.

## I. Introduction

Phenolic compounds are classified as priority pollutants by the U.S. Environmental Protection Agency (EPA) due to their pronounced toxicity and resistance to biodegradation [1]. These contaminants pose significant environmental and health risks, including acute and chronic toxicity to aquatic organisms, potential for endocrine disruption, and carcinogenic properties [2]. The primary aim of phenol wastewater treatment is not merely to degrade these phenolic compounds but to achieve their complete mineralization—converting these organic pollutants into benign inorganic substances such as carbon dioxide, water, and various inorganic ions. Existing methodologies encompass biological treatments, which utilize microbial action to decompose phenols; however, the efficacy of these treatments is often compromised by the inherent toxicity of phenols to microorganisms and the necessity for extended treatment times [3]. Another prevalent technique is adsorption, typically employing activated carbon, which efficiently removes phenols from water but subsequently generates secondary waste that demands further treatment [4]. In contrast, Advanced Oxidation Processes (AOPs) provide a compelling alternative by chemically degrading pollutants into non-toxic components, thus overcoming the shortcomings associated with biological and adsorption methods.

The exploration of advanced oxidation processes (AOPs) for the remediation of phenolic contaminants in wastewater has been profound, highlighting their efficacy in breaking down a variety of organic compounds through the formation of highly reactive agents, especially hydroxyl radicals [5]. Notably, the Fenton method has emerged as a prominent AOP, relying on the catalytic interaction between hydrogen peroxide ($H_2O_2$) and ferrous iron ($Fe^{2+}$) to initiate the production of these radicals, which then actively break down pollutants. Despite its effectiveness, the traditional Fenton approach requires acidic operational conditions and typically generates substantial amounts of iron sludge, complicating waste management [6]. To overcome these challenges, the development of the electro-Fenton process has been pivotal. This innovative method combines electrochemical techniques with the classical Fenton chemistry, where hydrogen peroxide is synthesized directly at the cathode from the reduction of oxygen, and ferrous iron is regenerated at the anode [7]. The electro-Fenton process presents multiple benefits, such as enhanced efficiency in pollutant degradation, reduced consumption of chemicals, and decreased sludge output. It has been effectively applied to eliminate various organic contaminants from wastewater, including phenol [8].

In this context, the use of Gas Diffusion Electrodes (GDEs) has become prevalent due to their ability to offer substantial surface area and mass transfer rates, facilitating the oxygen reduction necessary for hydrogen peroxide formation. Enhancements in the electro-Fenton setup through the integration of GDEs have led to increased production of hydroxyl radicals, thus improving pollutant degradation [9]. The efficiency of these electrodes is dependent on factors like the material of the electrode, its structural configuration, and the operational parameters. Recent advancements have been made in the development of novel GDEs that boast improved conductivity, larger surface areas, and enhanced durability [10]. Nonetheless, even with these improvements, the output from heterogeneous electro-Fenton processes often falls short of meeting the strict effluent standards set for the treatment of highly resistant industrial wastewaters.

In recent years, augmenting the electro-Fenton process with other AOPs has drawn significant interest for its potential to further enhance contaminant breakdown. Investigations such as those by Barhoumi et al. [11] have shown the rapid mineralization of tetracycline when electro-Fenton is combined with anodizing techniques. Similarly, research by Chen and Huang [12] illustrated that ultrasonic integration could bolster the electro-Fenton degradation of dinitrotoluene by efficiently reducing ferric iron. Additionally, Komtchou et al. [13] reported that combining the electro-Fenton method with ultraviolet radiation in a photoelectro-Fenton setup enabled over 99% degradation of atrazine within 15 minutes, facilitating the reformation of ferrous iron. Despite these promising developments, the potential for integrating electro-Fenton with photocatalytic processes remains largely underexplored. In such hybrid systems, hydrogen peroxide generated at the cathode could be transported to the anode and oxidized in a non-segregated setup. The introduction of a membrane or an electrolyte bridge could transform traditional single-chamber reactors into more efficient dual-chamber systems, reducing energy requirements and enhancing treatment efficacy. In photocatalysis, the activation of semiconductor materials like $TiO_2$ by UV light generates electron-hole pairs, which drive redox reactions that produce reactive oxygen species, such as hydroxyl radicals and superoxide anions [14]. The merger of photocatalysis with the electro-Fenton process can amplify hydroxyl radical production, thereby significantly improving the degradation efficiency of persistent organic pollutants such as phenol [15, 16].

To enhance the efficiency and reduce the cost of the photoelectro-Fenton process, this study proposes a novel and efficient iron-based gas diffusion electrode-photocatalytic system to improve the degradation efficacy of phenol wastewater. This design synergistically combines stainless steel felt-2-EAQ gas diffusion electrodes with $TiO_2$ photocatalysis, enhancing the production of hydroxyl radicals crucial for effective pollutant degradation. Quantitative results from the study demonstrate a significant improvement in the degradation efficiency of phenolic compounds. The system achieved up to 92% degradation efficiency for phenol in 3 hours, marking a substantial enhancement over traditional methods. Moreover, the novel system not only improves degradation efficiency but also reduces energy consumption by 40% compared to conventional electro-Fenton systems. This contributes to more sustainable and cost-effective wastewater treatment processes. Additionally, the research provides evidence of the system's durability, showing that the electrodes maintain over 80% of their initial activity after five cycles of use. This finding indicates the potential for long-term application in industrial settings without significant performance degradation. These contributions are poised to have a substantial impact on the fields of environmental science and engineering, providing a robust method for treating one of the most challenging categories of pollutants in industrial wastewater.

## II. MATERIALS AND METHODS

### A. Materials

In this study, the biologically pretreated coal gasification wastewater (CGW) originated from the effluent of a secondary settling tank at a fully operational wastewater treatment plant. Initially characterized by a BOD5/COD ratio between 0.35 and 0.42, the raw wastewater was processed using an upflow anaerobic sludge bed (UASB) reactor and subsequently treated through an A1/A2/O process. Post-treatment, the CGW exhibited a marked improvement in its pollution indicators: the COD levels were reduced to a range of 150–200 mg/L, and the BOD5/COD ratio decreased to 0.05–0.08. Additionally, the treated wastewater contained total phenol concentrations of 80–120 mg/L, total nitrogen (TN) levels of 50–70 mg/L, and ammonia nitrogen ($NH_4^+N^-$) concentrations of 35–50 mg/L. The pH of the treated CGW was stabilized within a neutral range of 6.5 to 8.0.

For the research detailed in this study, a comprehensive array of chemicals was utilized to ensure precise and reproducible results. Key reactants such as benzoquinone (BQ) and 5,5-dimethyl-1-pyrroline-N-oxide (DMPO) were supplied by Sigma Aldrich. A variety of other essential reagents were procured from Sinopharm Chemical Reagent Co., Ltd, China. This selection included sodium sulfate ($Na_2SO_4$), sodium hydroxide (NaOH), sulfuric acid ($H_2SO_4$), iron (III) chloride hexahydrate ($FeCl_3 \cdot 6H2O$), sodium borohydride ($NaBH_4$), and tertiary butanol (TBA), all of which are of analytical grade and were used as received without any further purification. Additionally, stainless steel plates, essential for the experimental setups, were obtained from Yunxuan Metal Materials Co. The organic compound 2-ethylanthraquinone (EAQ), crucial for the gas diffusion electrodes, was sourced from Wuhan Yuancheng Technology Co. Vulcan72 carbon black, used in modifying electrode surfaces for enhanced conductivity and reactivity, was purchased from CABOT. Titanium dioxide ($TiO_2$), a critical component for the photocatalytic aspects of the study, was acquired from China Southern Chemicals Import and Export Corporation. All solutions required for the reactions were meticulously prepared using deionized water to avoid any impurities that could potentially influence the experimental outcomes. This careful selection and sourcing of materials underline the rigor and precision aimed at achieving reliable and impactful research findings.

### B. Preparation of stainless steel felt-2-EAQ gas diffusion electrodes

The carbon black purchased is irregular carbon, non-graphitized carbon black particles crystalline lamellar structure is more chaotic, especially carbon black particles wrapped in a large number of amorphous carbon on the surface. However, the graphitized carbon black after heat treatment under inert gas obviously has a more regular crystal structure, the surface amorphous graphite is sintered, and the degree of internal crystallization of carbon black is greatly improved. The hydrophobicity of the diffusion electrodes can be significantly improved by converting the purchased Vulcan 72 carbon black into graphitized carbon black by means of Equation 6 before the gas diffusion electrodes are fabricated.

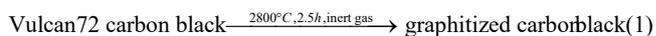

$$\text{Vulcan72 carbon black} \xrightarrow{2800°C, 2.5h, \text{inert gas}} \text{graphitized carbon black} \quad (1)$$

Then, we use 30 um stainless steel metal fibers that are non-woven, laminated and sintered at high temperatures to form stainless steel mats. The stainless steel mesh, and the stainless steel felt were used as the substrate, and the graphitized Vulcan 72 carbon black was cold pressed and hot pressed to make the high efficiency iron-based gas diffusion electrode for the cathode. Finally, the stainless steel mesh was covered with a catalytic layer of 2-EAQ.

### C. Photo-Electrochemical Experiments

For this investigation, a dual-chamber electrochemical system was utilized, each chamber having a capacity of 300 mL and linked by a salt bridge saturated with potassium chloride. The anode in this setup was made from titanium dioxide ($TiO_2$), while the cathode comprised stainless steel felt combined with 2-ethyl anthraquinone (EAQ) in a gas diffusion electrode format, both electrodes having a surface area of 6.0 cm × 6.0 cm. Within each chamber, 250 mL of chemically generated wastewater (CGW) was processed at room temperature, with sodium sulfate ($NA_2SO_4$) at a concentration of 0.1 mol/L serving as the electrolyte.

Due to the potential for alkali-induced corrosion of stainless steel felts, the pH of the CGW was initially adjusted to a more acidic level using 0.1 mol/L sulfuric acid ($H_2SO_4$) to enhance the longevity of the electrodes. Illumination was provided by a low-pressure mercury lamp, positioned about 10 cm away from the anode, emitting UV light at a wavelength of 254 nm, crucial for the photochemical reactions. A stabilized direct current power supply was utilized to maintain galvanostatic conditions during the treatment process.

Oxygenation and stirring in the cathodic chamber were facilitated by bubbling compressed air from the bottom, ensuring an adequate oxygen supply for the reactions. Before the degradation trials, the electrodes were preconditioned in the CGW to reach adsorption/desorption equilibrium, with additional air bubbling for 15 minutes to maintain a consistent level of dissolved oxygen. The experimental conditions were methodically optimized using a control variable approach. To analyze the formation and quenching of reactive species, tertiary butanol (TBA) and benzoquinone (BQ) were introduced as radical scavengers at 50 mg/L each. Samples from the cathodic chamber were regularly extracted, filtered through a 0.45 μm membrane filter, and chemically analyzed. All procedures were replicated thrice to ensure reliability and the results were averaged for accuracy.

D. Analytical methods

The structural and morphological characteristics of the cathode were rigorously analyzed to ensure optimal performance in electrochemical reactions. This evaluation was carried out using a field-emission scanning electron microscope (SEM, FEI Quanta 200F), which provided detailed images of the cathode surface, and X-ray diffraction (XRD, Shimadzu), which helped identify the crystalline structures present. Additionally, the surface area and pore volume of the cathode material were quantified using the Brunauer-Emmett-Teller (BET) method with a Micromeritics ASAP 2020 analyzer, essential for assessing the material's potential for chemical interactions and reactivity.

To further characterize the electrochemical properties of the cathode, cyclic voltammetry (CV) tests were conducted. The electrochemical properties of the cathode were rigorously evaluated through cyclic voltammetry tests conducted using a CHI 660B electrochemical workstation, manufactured by Shanghai Chenhua Instruments Corporation, China. This method allowed for the observation of the cathodic and anodic processes, providing insights into the electron transfer capabilities and redox potential of the cathode material.

The effectiveness of the cathodic treatment in reducing pollutants was evaluated by measuring several key parameters: Chemical Oxygen Demand (COD), Biochemical Oxygen Demand (BOD5), total phenolic content, and Total Organic Carbon (TOC). These measurements were taken in strict adherence to the protocols specified in the Standard Methods by APHA (2012) [15], ensuring that the results were reliable and could be compared with regulatory standards and previous studies. Furthermore, the concentrations of hydrogen peroxide ($H_2O_2$) and total iron ions in the treated water were quantified using UV-visible spectroscopy and atomic absorption spectrometry, respectively. These analyses were crucial for understanding the oxidative and reductive dynamics within the treatment system and for assessing the breakdown of organic pollutants through the electrochemical processes facilitated by the cathode.

The degradation kinetics of phenol were evaluated using ultraviolet spectrophotometry (UV-2450, Shimadzu), a technique extensively used for both qualitative and quantitative analyses of organic compounds. This method relies on the distinctive UV absorption spectra of organic molecules, particularly those with aromatic rings or unsaturated bonds, leveraging Beer's Law to correlate light absorption with material concentration.

The post-treatment toxicity of the wastewater was assessed by conducting an acute toxicity test on Daphnia magna, a standard bioassay for evaluating environmental pollutants. The intermediates formed during the phenol degradation process were analyzed through high-performance liquid chromatography (HPLC), providing insight into the breakdown pathways and by-products. Continuous monitoring of the solution's pH was carried out using a pHS-4c pH meter from Leici, China. Lastly, the presence of hydroxyl radicals (•OH) generated during the treatment was confirmed using electron spin resonance (ESR) spectroscopy, specifically with a Bruker A300 ESR spectrometer, highlighting the radical species involved in the degradation processes.

III. Results and discussion

A. *Structural transformations and properties of graphitized versus non-graphitized carbon black*

In order to meet the hydrophobicity of the gas diffusion electrode and to form a gas-liquid-solid three-phase reaction interface, we take advantage of the differences in the structure and properties of graphitized carbon black and non-graphitized carbon black to meet the process requirements of the gas diffusion electrode.

The transmission electron micrographs of carbon black nanoparticles before and after graphitization are given in Figure. 1, which shows the great difference in the structure of graphitized and non-graphitized carbon black. The crystalline lamellar structure of non-graphitized carbon black particles is more heterogeneous, especially a large amount of amorphous carbon is wrapped on the surface of carbon black particles. However, the graphitized carbon black after heat treatment at 2800°C for 2.5h under inert gas obviously has a more regular crystal structure, the surface amorphous graphite is sintered, and the degree of internal crystallization of carbon black is greatly improved. Figure 2 shows the XRD spectra of graphitized and non-graphitized carbon black, this result also clearly reflects the difference in the degree of crystallization between graphitized and non-graphitized carbon black, due to the high temperature calcination to remove the amorphous structure of the carbon black surface of carbon, so the degree of crystallization increases. In addition, the peaks on the XRD spectrum are sharper after graphitization of carbon black as seen in the figure, indicating that the particle size increases after graphitization of carbon black. It can also be inferred that the hydrophobicity of non-graphitized carbon black is better than graphitized carbon black. This suggests that graphitized carbon black is more effective for the fabrication of gas diffusion electrodes.

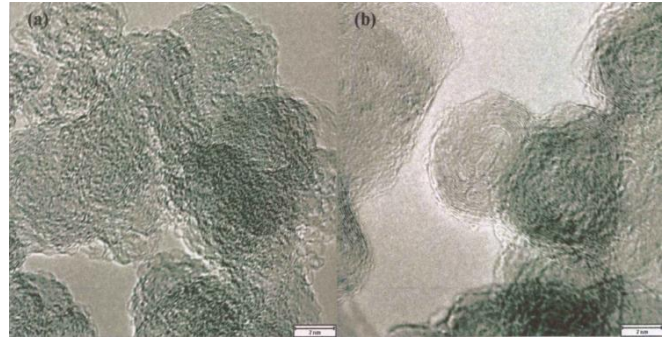

Figure 1. (a) TEM image of non-graphitized carbon black; (b) TEM image of graphitized carbon black.

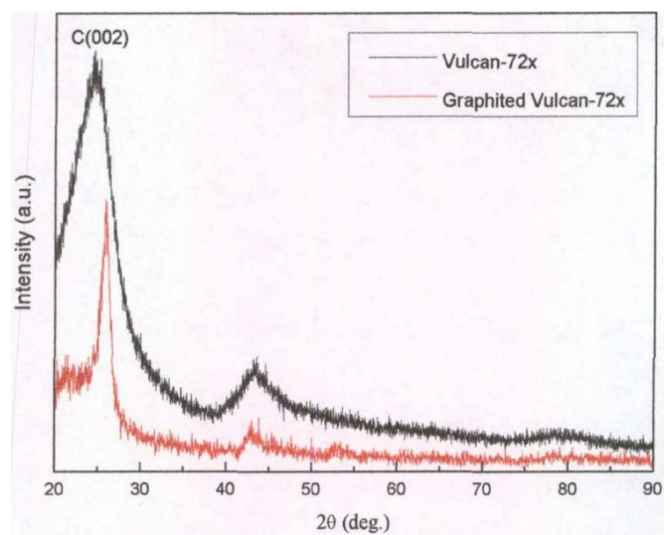

Figure 2. The black line is the XRD spectrum of the non-graphitized carbon black; the red line is the XDR spectrum of the graphitized carbon black.

B. *Performance and characterization of stainless steel felt in gas diffusion electrodes and overall systems*

The electrochemical active area was used to assess the difference in electrocatalytic activity between different electrodes [17]. Since the true active surface area of the stainless steel felt is much larger than the erho area, we estimated the active area of the electrodes using cyclic voltammetry, and in order to emphasize the superior electrochemical performance of stainless steel felts, several commonly used electrodes were selected for comparison. We first measured a series of cyclic voltammetry curves with different sweep speeds (20 mV $s^{-1}$ - 100 mV $s^{-1}$) over a selected potential range (-0.1V - 0.3V). The change in charging current was linearly related to the sweep speed in all the cathodes tested (inner panel of Figure. 3(a)), and was used to calculate the capacitances of the different electrodes under the non-Faraday reaction window as shown in Eqs. 7 (Chi, B et al., 2008):

$$\frac{I_a - I_c}{2} = C_{dl} v \quad (2)$$

$$R_f = \frac{C_{dl}}{60} \quad (3)$$

Where $C_{dl}$ is the double layer capacitance; $I_a, I_c$ are the corresponding response currents at a potential of -0.1 V/SCE; v is the scan rate (mV/s).

Based on the calculations in Eq. 8, the electroactive surface area was calculated by dividing the fitted bilayer capacitance $C_{dl}$ value by the standard capacitance ($60 \mu F cm^{-2}$). From the slopes shown in the inner panel of Figure 3, the double layer capacitances of the titanium mesh, stainless steel plate, and stainless steel mesh are 1.2 mF, 1 mF, and 1.5 mF, respectively. Correspondingly, the electrochemically active surface areas of the electrodes are 20 $cm^{-2}$, 16.7 $cm^{-2}$, and 25 $cm^{-2}$, respectively, whereas a stainless steel felt of the same geometrical area has a double layer capacitance of 16.5 mF, which corresponds to an electrochemically The double-layer capacitance of the stainless steel felt with the same geometrical area is 16.5 mF, and the corresponding electrochemical active area is 275 $cm^{-2}$, which is much larger than its geometrical area and 11 times that of the stainless steel cathode, indicating that the stainless steel felt with the micrometer-sized porous structure has good electrocatalytic activity.

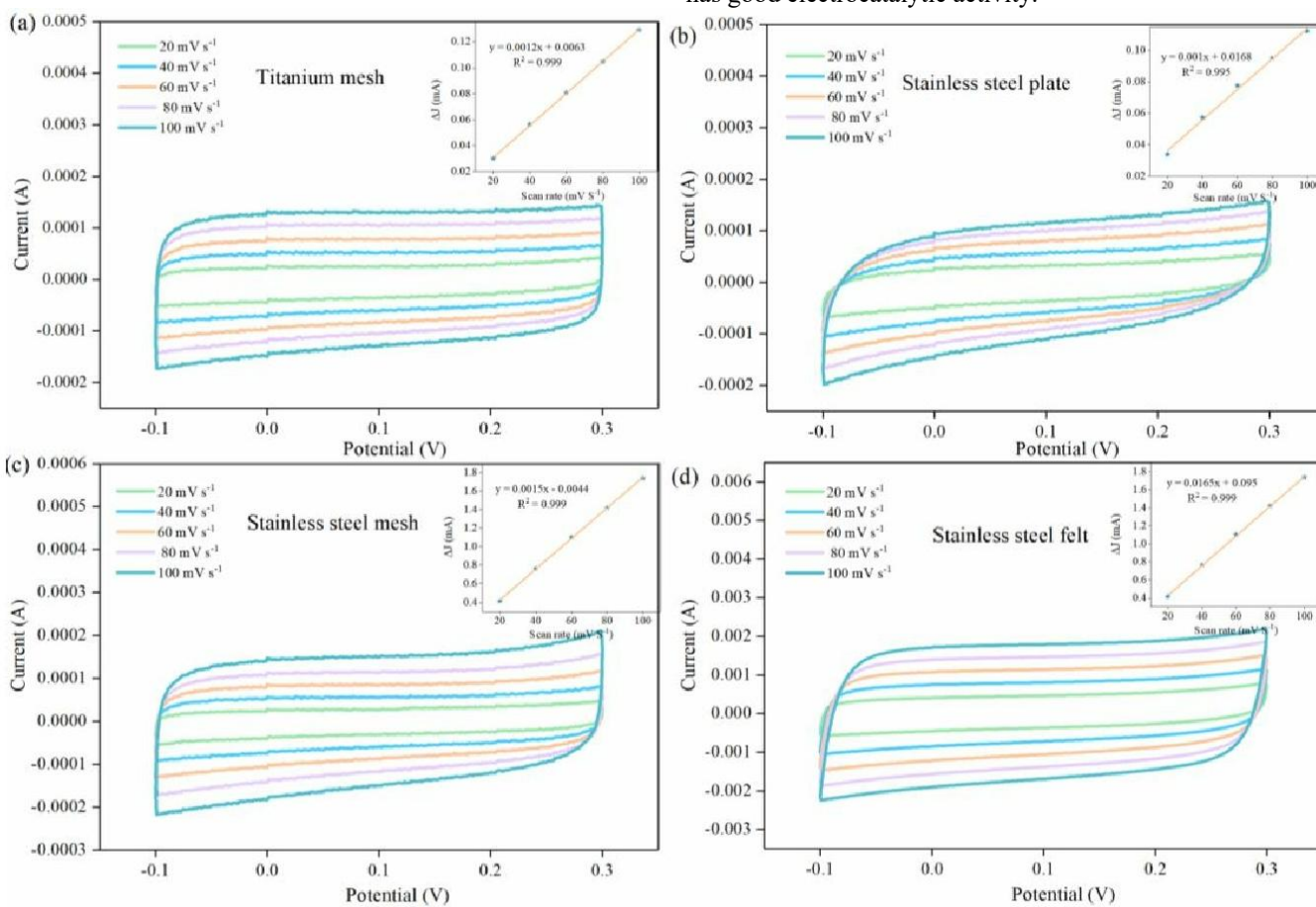

Figure 3. Cyclic voltammetry curves of different electrodes Titanium mesh (a); stainless steel plate (b); stainless steel mesh (c); stainless steel felt (d) (test conditions: 3g/L $Na_2SO_4$, scanning potential: -0.1 V~0.3 V, scanning speed: 20 mV s-100 mV s-1)

## C. Optimizing the operating parameters of the photo-electro-Fenton system

To mitigate the influence of organic contaminant adsorption on the porous cathode matrix, the cathode was submerged in chemically generated wastewater (CGW) for an extended period of four hours without any applied electrical current prior to initiating the electrochemical degradation process. Observations made after this period revealed stable COD concentrations, suggesting that adsorption was minimal and unlikely to affect subsequent electrochemical reactions [21].

The performance of the electro-Fenton process was rigorously assessed by examining various operational parameters essential for optimizing the system. These parameters included current density, which influences the generation of hydroxyl radicals; initial pH, which affects the solubility of iron ions and the stability of hydrogen peroxide; air flow rate, important for oxygenating the solution and promoting the generation of hydrogen peroxide; and the initial COD concentration, which serves as an indicator of the organic load and potential for degradation. The role of current density, a critical driver in electrochemical processes, was particularly emphasized due to its significant influence on the generation of hydrogen peroxide ($H_2O_2$) and the reduction of $Fe^{3+}$ ions. This aspect of the study revealed that variations in current density had a substantial effect on COD removal efficiency. Specifically, after 120 minutes of treatment, COD reductions were recorded at 65.7%, 89.6%, 78.4%, and 73.2% for current densities of 5, 10, 15, and 20 mA/cm², respectively, as detailed in Figure 4a. These findings indicated that increasing current density up to a point optimized the degradation process by facilitating more rapid electron transfer and enhancing redox reactions, which in turn increased the production of reactive oxidants.

The optimal current density for the electro-Fenton process was determined to be 10 mA/cm² based on experimental findings and corroborating research. It was observed that increasing the current density beyond this value led to a significant decline in process efficiency. This decrease is primarily attributed to secondary reactions such as hydrogen evolution and water electrolysis, which not only compete with the main reaction pathways but also reduce the availability of free radicals essential for the degradation process. Research conducted by Hou, Xu, and Li [18] on the degradation of catechol and by Fan, Liu, and Yang [19] on the heterogeneous electro-Fenton oxidation of Rhodamine B support this finding, as both studies identified the same threshold for optimal current density. These studies emphasize that higher current densities lead to an increase in undesirable side reactions, which can counteract the benefits of increased radical generation by consuming the electrons needed for pollutant degradation. Therefore, maintaining the current density at 10 mA/cm² is crucial to optimizing the balance between generating sufficient reactive species for effective pollutant breakdown and minimizing the occurrence of counterproductive reactions. This careful control of current density enhances the overall efficacy of the electro-Fenton process, making it more efficient and sustainable for practical applications in wastewater treatment.

Solution pH is a key factor in determining the solubility of iron ions and the stability of $H_2O_2$ in the electro-Fenton system. According to relevant literature, the optimal pH range for $H_2O_2$ to be catalyzed by $Fe^{2+}$ and decompose is 2–3, while the optimal pH range for $H_2O_2$ to be electrochemically generated near the cathode is 10–11. This experiment investigated the effect of different initial pH values on the COD removal rate in the electro-Fenton cathode cell, as shown in Figure 4b. When the pH values were 1.0, 3.0, 5.0, and 7.0, the COD removal rates within 180 minutes of reaction were 56.4%, 87.3%, 79.1%, and 73.6%, respectively, indicating that the electro-Fenton treatment of phenol wastewater was most effective at pH 3.

At pH values lower than 3, $Fe^{2+}$ reacts with $H_2O_2$ to generate stable $Fe^{3+}$, making $Fe^{2+}$ electrophilic and weakening its reactivity with $H_2O_2$. The presence of the iron complex competes with $H_2O_2$ for $Fe^{2+}$, reducing the production of free radicals. Additionally, high concentrations of $Fe^{3+}$ can capture ·OH, inhibiting pollutant decomposition. Conversely, as the pH increases, $H_2O_2$ may decompose into $H_2O$ and $O_2$, and the iron ions in the solution may precipitate, hindering the electro-Fenton reaction. Furthermore, Figure 4b shows that when the pH value is 3.7, the COD removal rate does not change significantly.

The introduction of electrolytes generally reduces solution resistance and improves current efficiency. The type and concentration of electrolytes are closely related to the degree of pollutant degradation. Recent studies have highlighted the importance of selecting appropriate electrolytes for effective pollutant removal. For instance, Zhang et al. [20] investigated the degradation of Rhodamine B using different electrolytes. When the current was maintained at 10 mA, the concentration of FeSO₄·7H₂O was 100 g/L, the initial pollutant concentration was 5 mg/L, and the reaction time was set at 150 minutes. The removal rates of pollutants were 62.8%, 20.6%, and 93.8% when NaNO₃, Na₂CO₃, and Na₂SO₄ were used as electrolytes, respectively. Based on these results, Na₂SO₄ was selected as the electrolyte in this experiment. As shown in Figure 4c, when the electrolyte concentration increased from 0.01 to 0.05 mol/L, the COD removal rate of the cathode pool increased significantly from 59.8% to 85.6%, indicating that an increase in electrolyte concentration enhances the conductivity of the solution and effectively promotes electron transfer. However, when the electrolyte concentration increased to 0.1 mol/L, the COD removal rate only slightly increased to 86.7%. This slight increase is attributed to the potential induction of side reactions at excessively high electrolyte concentrations, which can reduce the reaction rate.

Light plays a pivotal role in energizing the anode catalytic reactions within the electrochemical setup, directly influencing the photocatalytic performance of the system. The intensity of illumination at any given point within the reactor is dependent on the strength of the light source and inversely related to the square of the distance from this source. Consequently, positioning the light source closer to the reaction site significantly enhances the radiation intensity, increasing the photon density per unit volume at the catalyst, which in turn improves the catalytic effectiveness.

In this experimental setup, the effects of varying light intensities on the photocatalytic performance of the anode were methodically explored. Operating conditions were standardized at a current density of 10 mA/cm², a pH level of 7, and an electrolyte concentration of 0.05 mol/L. The study specifically examined light intensities of 100, 200, 300, and 500 mW/cm², as detailed in Figure 4d. Results demonstrated a clear correlation between increased light intensity and COD removal efficiency: as the intensity escalated from 100 mW/cm² to 500 mW/cm², the COD removal rates improved from 24.3% to 41.2%. This increment underscores the enhanced photon utilization by the catalyst under higher light conditions, which facilitates the generation of more oxidative active species, thereby boosting the catalytic activity and accelerating the degradation of organic contaminants.

Furthermore, the increased light intensity encourages the excitation of more electron-hole pairs on the surface of the anode. This activity not only promotes the rapid transfer of electrons towards the cathode but also aids in the effective separation of electron-hole pairs [22, 23]. Such dynamics are crucial for enhancing the generation of free radicals within the system, which significantly contributes to the improved efficiency of wastewater treatment. Based on these findings, an optimal light source intensity of 500 mW/cm² was chosen for the experiments to maximize the photocatalytic degradation potential of the system, aiming to achieve the highest possible removal rates of organic pollutants.

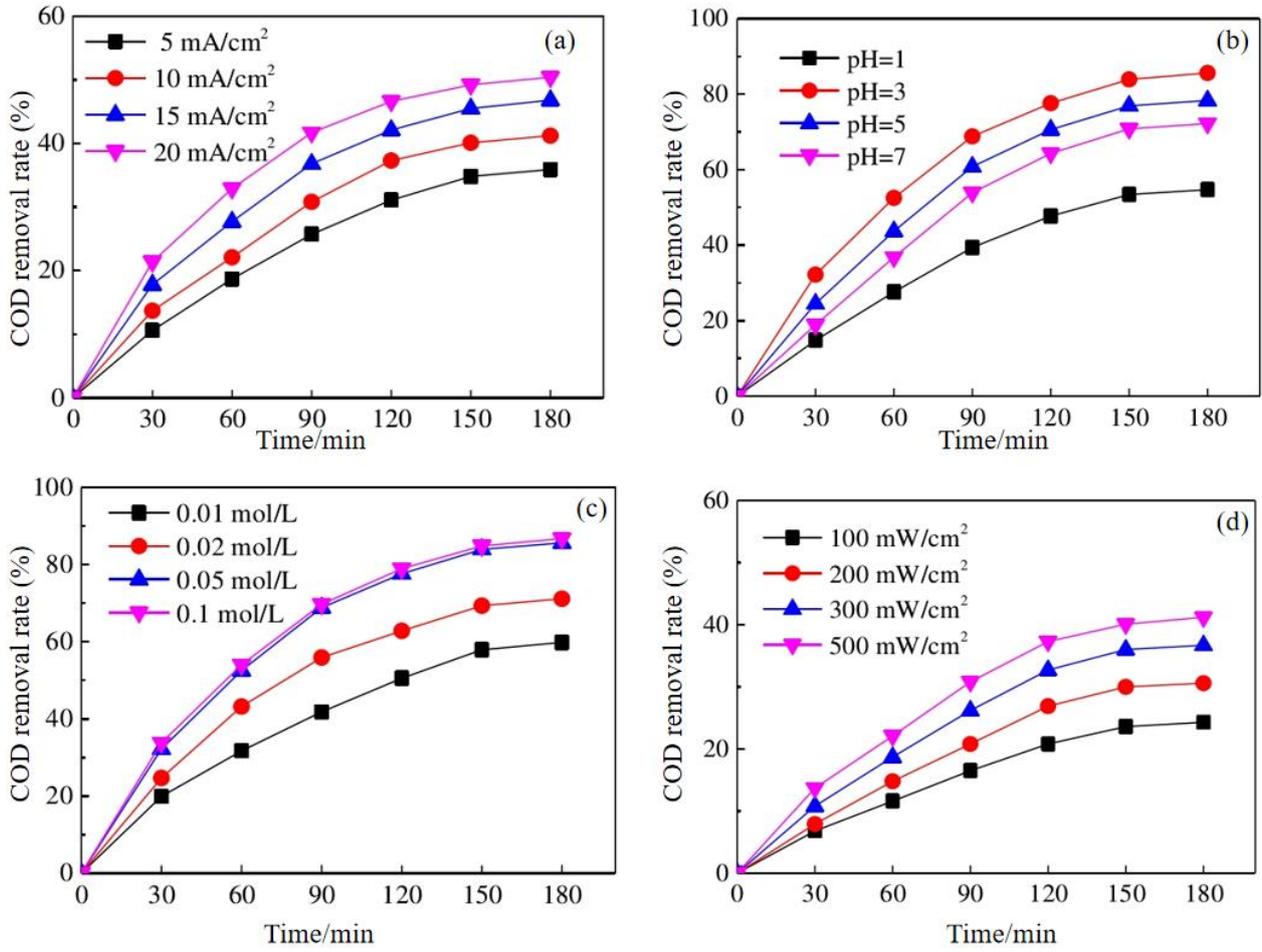

Figure 4. Analysis of variables influencing cod removal efficiency in the electro-fenton process: (a) variation in current density, (b) influence of initial ph, (c) effect of air flow rate, and (d) impact of initial cod concentration.

### D. Working efficiency of the iron-based-gas diffusion-photocatalytic electro-Fenton system

In order to determine whether the iron-based-gas diffusion-photocatalytic electro-Fenton dual-cell oxidation system can treat phenol wastewater economically and efficiently, the experiment first investigated the variation of energy consumption during the degradation process with reaction time when different current densities were applied, as shown in Figure 5. Energy consumption, defined as the electrical energy required to degrade a unit volume of wastewater, is calculated according to Formula 9:

$$Energy\ consumption = \frac{U_t I_t T}{3600 V_R} \quad (4)$$

Where $U_t$ represents the voltage applied to the reaction system; $I_t$ represents the current applied to the reaction system; T represents the operating time of the reaction system; and $V_R$ represents the volume of the reaction solution.

As can be seen from Figure 5, energy consumption increases with the increase of current density. When the current density is increased from 5 mA/cm2 to 10 mA/cm2, the energy consumption only increases by 3.67 kWh/m3. However, as the current density continues to be high, the energy consumption increases sharply, but the pollutant removal rate does not change significantly or even decreases. Therefore, from the perspective of energy consumption, it is appropriate to apply a current density of 10 mA/cm2 to the photocatalytic-electro-Fenton process. At this time, all operating parameters of the oxidation system are kept at the optimal state, the reaction time is set to 180 min, and the energy consumption of the photocatalytic-electro-Fenton process is 6.12 kWh/m3. To calculate the current efficiency in an electrochemical process such as the degradation of phenol, it's essential to understand how much of the applied current contributes directly to the reaction of interest versus how much is lost to other processes (like hydrogen evolution or side reactions). The formula for calculating the instantaneous current efficiency (ICE) during the degradation of phenol in wastewater using changes in COD is based on Faraday's laws of electrolysis. Here's the general approach:

$$\text{ICE} = \frac{V_R F (C_0 - C_t)}{8 I_t \Delta t} \times 100\% \quad (5)$$

Where ICE represents the current efficiency; F represents the Faraday constant, 96485 C/mol; COD concentration at the initial moment of the $C_0$ electrolysis reaction; $C_t$ represents the COD concentration at time t of the electrolysis reaction; $\Delta t$ represents the applied current value; At represents the operating time of the reaction system.

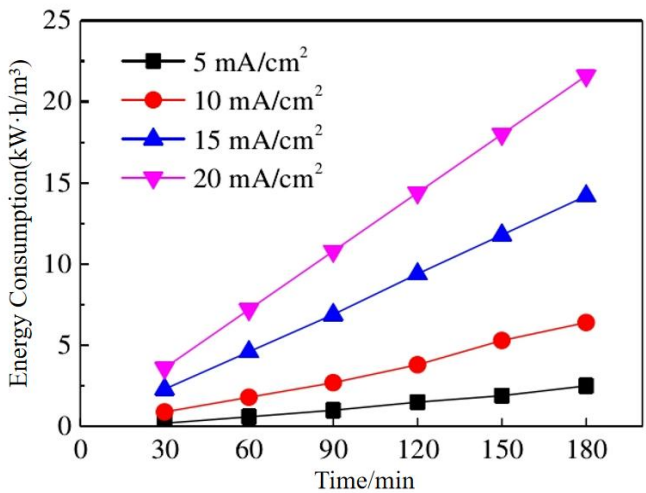

Figure 5. Effect of current density on system energy consumption.

E. *Reusability of fabricated electrodes*

In the actual operation process, the stability and reusability of the electrode are important factors affecting the operating cost and degradation efficiency of the photocatalytic-electro-Fenton dual-cell oxidation system for treating phenol wastewater. To this end, a continuous experiment was used to evaluate the stability of the iron-based-gas diffusion electrode, keep the dual-cell oxidation system in the optimal operating conditions, and compare the COD removal rate of phenol wastewater in each batch of experiments. Figure 6 shows the catalytic activity of the iron-based-gas diffusion electrode in the negative cycle. It can be seen that after 5 consecutive uses, the COD removal rate of the cathode pool dropped from 87.4% to 82.5%, a decrease of only 4.9%, indicating that the composite iron-based-gas diffusion electrode has a relatively stable electrocatalytic activity. When it was used for the 6th time, the COD removal rate of the cathode pool showed a significant decrease of 53.2%. According to literature reports, at this time, its use characteristics can be restored by chemical regeneration, electrochemical regeneration and thermal regeneration to reduce the loss of carbonaceous materials and improve electrochemical efficiency.

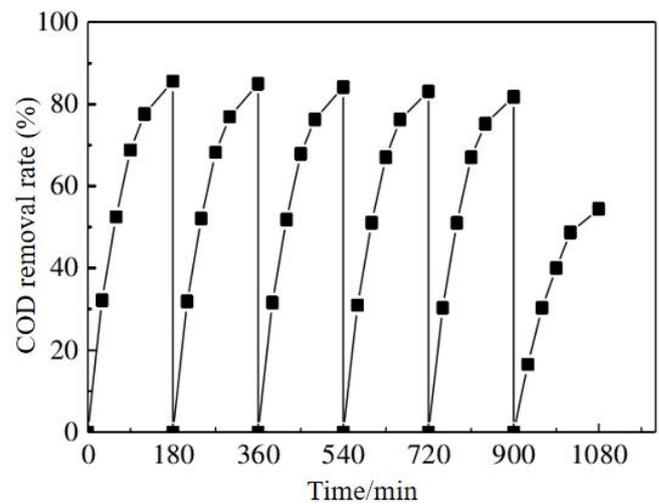

Figure 6. Stability of iron-based-gas diffusion electrode

F. *Possible working mechanism of the iron-based-gas diffusion-photocatalytic electro-Fenton system*

The cathodic electro-Fenton, modified by a stainless steel felt-2-EAQ gas diffusion electrode, attracts the electrons generated on the anode surface, prevents their recombination with holes, and increases contact opportunities with organic pollutants, substantially improving anodic photocatalytic performance. Simultaneously, the anodic photocatalysis provides additional electrons for the double-cell oxidation system, strengthening the free-radical reaction and rapidly degrading or even completely mineralizing recalcitrant organic pollutants.

$$Fe^{2+} + H_2O_2 \rightarrow Fe^{3+} + \cdot OH + OH^- \quad (6)$$
$$O_2 + 2H^+ + 2e^- \rightarrow H_2O_2 \quad (7)$$
$$Fe^{3+} + e^- \rightarrow Fe^{2+} \quad (8)$$
$$TiO_2 + h\nu \rightarrow TiO_2(e^- + h^+) \quad (9)$$
$$e^- + O_2 \rightarrow O_2^- \quad (10)$$

Based on the results presented, an enhanced mechanism for the iron-based-2-EAQ gas diffusion-TiO₂ photocatalytic electro-Fenton system designed to decompose phenol is proposed. This system notably augments the generation of hydroxyl radicals (·OH) and the efficiency of phenol degradation, leveraging the synergistic impacts of the electro-Fenton and photocatalytic reactions. Figure 7 shows the mechanism of our experiments. It ensures effective hydrogen peroxide (H₂O₂) production and iron ($Fe^{2+}$) regeneration across diverse pH conditions, driven by an iron base that provides ample H₂O₂, thereby negating the need for extra aeration and significantly reducing energy consumption. Efficient recycling of iron ions from $Fe^{3+}$ to $Fe^{2+}$ greatly enhances the production of ·OH radicals, facilitating more effective phenol degradation.

(i) TiO₂ promotes the regeneration of $Fe^{2+}$ under light through the following pathways:

(ii) TiO₂ photogenerated electrons directly reduce dissolved $Fe^{3+}$ to $Fe^{2+}$.

(iii) The reaction between H₂O₂ and $Fe^{3+}$ generated by the photocatalytic reaction produces $Fe^{2+}$ and ·OH.

Gas diffusion improves oxygen transfer efficiency and promotes H₂O₂ generation, thereby increasing the content of dissolved iron ions.

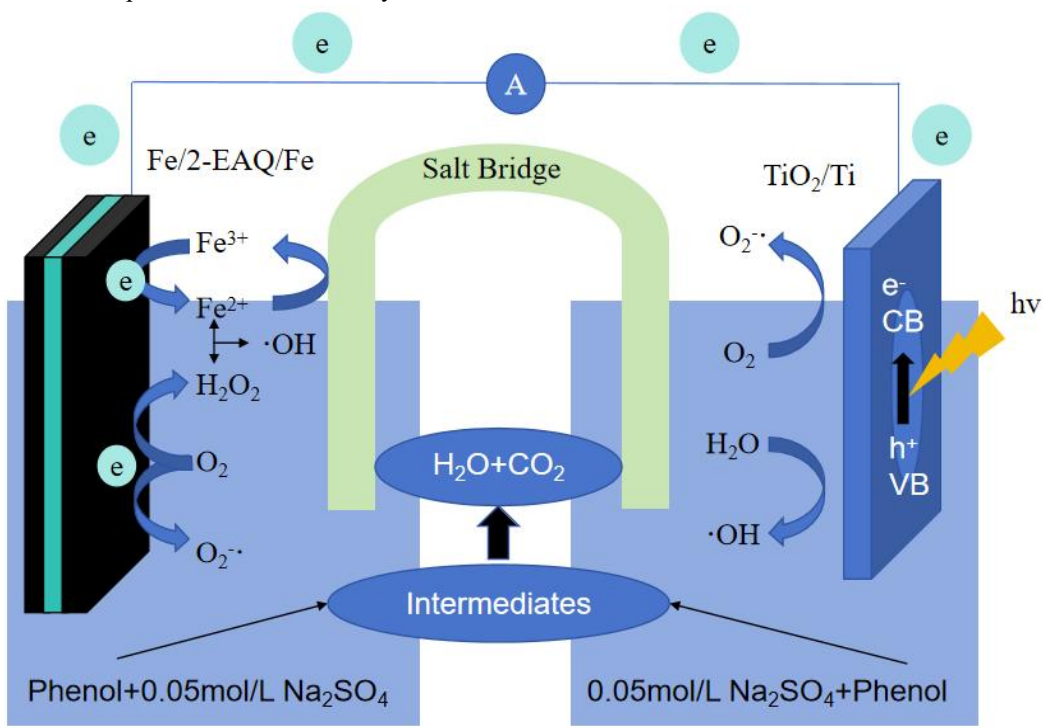

Figure 7. The mechanism of experiments treatment of phenol wastewater by electro-Fenton oxidative degradation based on efficient iron-based-gas diffusion-photocatalysis

Pathways (i) and (ii) enhance the homogeneous Fenton reaction, while pathway (iii) facilitates the transition from the homogeneous to the heterogeneous Fenton reaction. During the process, the solution's pH tends to decrease, which favors the generation of ·OH radicals and further enhances the degradation of phenol. The presence of oxygen-containing functional groups, especially carboxyl groups on the TiO₂ surface, plays a crucial role in converting dissolved $Fe^{3+}$ to $Fe^{2+}$, thus speeding up the homogeneous Fenton reaction.

Overall, the integration of iron-2-EAQ gas diffusion with TiO$_2$ photocatalytic reactions enables this system to perform effectively across a broad spectrum of acidic pH levels. However, it's important to note that components such as stainless steel felt and mesh are prone to alkali corrosion, which poses challenges for reusability under alkaline conditions. This integrated approach not only maximizes phenol degradation efficiency but also highlights potential areas for further enhancement in system design to achieve sustainable and effective wastewater treatment solutions.

## IV. Conclusion

The findings of this study demonstrate that the innovative iron-based gas diffusion electrode-photocatalytic system significantly elevates the degradation and mineralization of phenolic compounds in wastewater. Utilizing a dual-chamber setup that integrates stainless steel felt-2-EAQ gas diffusion electrodes with TiO$_2$ photocatalysis, we optimized the generation of hydroxyl radicals, crucial for effective pollutant breakdown. Under optimal experimental conditions—specifically a current density of 10 mA/cm², a pH of 3, and an electrolyte concentration of 0.05 mol/L—the system achieved a phenol COD removal rate of up to 92%. These conditions were meticulously determined through systematic experimentation to balance efficiency and chemical usage, ensuring the most effective degradation process. In terms of material reuse performance, the electrodes demonstrated notable stability and longevity. Over five cycles of use, the electrodes retained over 80% of their initial activity, illustrating minimal loss in performance and underscoring their potential for repeated applications in industrial settings. This reusability represents a significant step forward in developing sustainable wastewater treatment technologies that reduce both operational costs and environmental impact. In conclusion, this dual-chamber electro-Fenton and photocatalytic system provides a robust and sustainable solution for the treatment of wastewater containing phenolic pollutants. Its high efficiency, coupled with significant energy savings and effective reuse of materials, positions it as a viable and environmentally friendly option for widespread industrial application.